\documentclass[10.5pt,a4paper,oneside]{extarticle}
\usepackage[T1]{fontenc} % package for character coding
\usepackage{geometry}
    \usepackage[english]{babel}
    \usepackage[affil-it]{authblk}
    \usepackage{etoolbox}
    \usepackage{lmodern}
    \usepackage{setspace}
    \usepackage{amssymb,amsthm,amsmath}
    \usepackage{tikz}
    \usepackage{listings}
    \usepackage{color}
    \usepackage{caption}
    \usepackage{placeins}
    \usepackage{booktabs}
    \usepackage{enumerate}
    \usepackage{graphicx}
    \usepackage{float}
    \usepackage{titlesec}
    \usepackage{mathptmx}
    \usepackage{anyfontsize}
    \usepackage{t1enc}
    \usepackage[utf8]{inputenc}
    \usepackage{fancyhdr}
    \usepackage{float,xspace}
    \usepackage{scalefnt}
    \usepackage[numbers]{natbib}
    \usepackage{mwe}
    \usepackage{afterpage}
    \usepackage[flushleft]{threeparttable}
    \usepackage{tabularx}
    \usepackage{booktabs}% nicer horizontal lines
    \usepackage{caption}% fix vertical spacing of table captions
   \usepackage[]{sidecap}
   \sidecaptionvpos{figure}{c} 
   \usepackage[font=small,labelfont=bf]{caption}
    \usepackage{siunitx}% align numbers at the decimal point
    \let\oldbibliography\thebibliography
    \renewcommand{\thebibliography}[1]{%
    \oldbibliography{#1}%
    \setlength{\itemsep}{0pt}%

}
    
\usepackage{soul} % to be able to use \hl{} command
\usepackage{tabulary}
\usepackage[colorlinks=true,linkcolor=blue]{hyperref} % added by AO

\usepackage{textcase}

\usepackage[colorinlistoftodos]{todonotes}

\renewcommand{\arraystretch}{1.5} % Table spacing, the default is 1.0

\titlespacing*{\section}
{0pt}{12pt}{0pt}
\titlespacing*{\subsection}
{0pt}{12pt}{0pt}
\titlespacing*{\subsubsection}
{0pt}{12pt}{0pt}
\titleformat{\section}
  {\huge\sffamily\Large\bfseries\color{black}}
  {\thesection}{1em}{}
    \titleformat{\subsection}[block]{\bfseries}{\thesubsection}{.5em}{}
    \titleformat{\subsubsection}[block]{\bfseries}{\thesubsubsection}{.5em}{}

\titleformat{\section}{\fontsize{12}{19}\bfseries}{\thesection}{1em}{}

\usepackage{mathptmx} % 'Times new roman'

\makeatletter

\patchcmd{\@maketitle}{\LARGE \@title}{\fontsize{14}{19.2}\selectfont\@title}{}{} % original: 18, 19.2
\makeatother
\title
{
	\vspace{-2cm}
	\begin{minipage}{\textwidth}

	\hspace{-20pt}%\includegraphics[width=9.5cm]{figures/ICA_logo.jpeg}
	\hspace{-65pt}%\includegraphics[width=10.2cm]{figures/ICA_text2022.png}
	\end{minipage}
	\includegraphics[width=16.5cm]{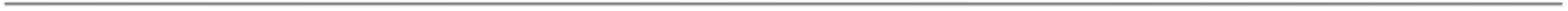}\\[0.5cm] \textbf{Improving spatial cues for hearables using a parameterized binaural CDR estimator}
	\author[ ]{Reza~Ghanavi$^{(1)}$, Craig~Jin$^{(2)}$}
  	\affil[(1)]{University of Sydney, Australia, reza.ghanavi@sydney.edu.au}
  	\affil[(2)]{University of Sydney, Australia, craig.jin@sydney.edu.au}
}
\date{}

\begin{document}
	\newgeometry{top=10cm}
    \afterpage{\aftergroup\restoregeometry}
% \begin{titlepage}
\clearpage
\setcounter{page}{1}
\maketitle
\thispagestyle{empty}
\fancypagestyle{empty}
{	
	\fancyhf{} \fancyfoot[R]
	{
		\vspace{-2cm}%\includegraphics[width=17cm]{figures/ICA_bar2022.png}
	}
}

\subsection*{\fontsize{10.5}{19.2}\uppercase{\textbf{Abstract}}}
{\fontsize{10.5}{60}\selectfont We investigate a speech enhancement method based on the binaural coherence-to-diffuse power ratio (CDR), which preserves auditory spatial cues for maskers and a broadside target. Conventional CDR estimators typically rely on a mathematical coherence model of the desired signal and/or diffuse noise field in their formulation, which may influence their accuracy in natural environments. This work proposes a new robust and parameterized directional binaural CDR estimator. The estimator is calculated in the time-frequency domain and is based on a geometrical interpretation of the spatial coherence function between the binaural microphone signals. The binaural performance of the new CDR estimator is compared with three state-of-the-art CDR estimators in cocktail-party-like environments and has shown improvements in terms of several objective speech quality metrics such as PESQ and SRMR. We also discuss the benefits of the parameterizable CDR estimator for varying sound environments and briefly reflect on several informal subjective evaluations using a low-latency real-time framework.}

\noindent{\\ \fontsize{11}{60}\selectfont Keywords: CDR estimation, binaural speech enhancement.} % at least 1 keyword is required (maximum of 5 keywords)

\fontdimen2\font=4pt

\section{\uppercase{Introduction}}
% \section{Introduction}

Speech enhancement and listening comfort improvement in multi-talker, noisy environments remains an active research area in binaural hearing and binaural signal processing for both hearables and hearing aids~\cite {arons1992review, ebata2003spatial, parande2017study, qian2018past}. Research has shown that exploiting the short-time spatial coherence estimate between two adjacent microphones is an effective way of calculating the gains required for spectral enhancement ~\cite {jeub2011blind, mccowan2003microphone, schwarz2015coherent, thiergart2012signal, thiergart2012spatial}. Among the model-based dereverberation methods, a limited number of them are proposed for binaural applications~\cite{jeub2010binaural}. An attractive feature of the binaural spatial coherence approach is that applying simple Wiener post-filtering on the short-time binaural spatial coherence preserves auditory spatial cues such as interaural time difference and interaural level difference for all sources when the target source is located directly ahead~\cite{jeub2010model,westermann2013binaural,schwarz2015coherent}. However, many previous coherence-based methods do not consider binaural processing per se but focus on speech enhancement.\par
In~\cite{jeub2011blind} a reverberation suppression method is introduced that estimates the coherent-to-diffuse energy ratio (CDR) for post-filtering gain calculation in a complex noise field. In particular, this work considers the geometry of the spatial coherence function in the complex plane for direct and diffuse sound components. Further research in~\cite{thiergart2012signal, thiergart2012spatial} shows that CDR-based dereverberation improves when the estimator accounts for the direction of arrival (DOA) of the direct signal and the phase of the complex-valued spatial coherence function. An issue with the aforementioned CDR estimators is that they rely on models of the coherence of the direct signal and/or diffuse noise field to estimate the CDR, and these models may not always match complex binaural noise fields in natural environments. A few numbers of the several heuristic CDR estimators proposed by Schwarz \emph{et. al.}~\cite{schwarz2015coherent} have shown greater robustness when compared with the CDR estimators introduced in \cite{jeub2011blind}, \cite{thiergart2012signal} and \cite{thiergart2012spatial}.  
As well, a more recent study by L\" {o}llmann \emph{et al.}~\cite{lollmann2021effective} estimates the CDR based on the effective rank of the covariance matrix of the input signals. Although this method does not require a coherence model for the signal and noise sound fields, it has the drawback of higher computational cost in real-time applications compared with the coherence-based method and produces less target to masking ratios in multi-talker environments due to its omnidirectionality.\par 
In this work, we propose a new robust directional CDR estimator derived from the complex-valued short time-frequency domain spatial coherence function between the observed binaural signals. The new CDR estimator requires neither a coherence model of the noise field nor an estimation of the room reverberation time. In this formulation, the target direction for the desired coherent signal is always broadside and straight-ahead. Hence, the symmetrical phase-magnitude response of the new estimator can be physically and psychoacoustically matched with any head shape and size to preserve natural binaural cues recorded by conchal microphones. In addition, the online adjustment of the new formula's real-valued parameter $S$ enables precise binaural dereverberation and denoising in a given acoustic environment.\par
This paper is organized as follows.
First, in Section~\ref{methods}, a novel parameterized CDR estimator is formulated and described. Further, in this section, the application of the new CDR estimator in reverberation suppression is illustrated, and the real-time implementation of the mentioned algorithm is briefly described.
 In Section~\ref{results}, the objective and perceived sound quality of the new binaural speech enhancement algorithm is compared with several state-of-the-art counterparts.

%%%
%%%
\section{\uppercase{methods}}
\label{methods}
\subsection{New CDR estimator}
\label{sec:cdr_calculation}
In this study, we consider the recording of a reverberated and/or noisy speech signal by two identical omnidirectional conchal microphones. We assume that the auto-power spectra of the microphone signals recorded for the broadside target are equal. In this case, the spatial coherence function, $\hat{\Gamma}_{l,r}(m,k)$, with the frame index $m$ and frequency $k$ for two binaural microphone signals, is expressed in the time-frequency domain as:
%%%%% EQUATION %%%%%
\begin{equation}\label{eq:5}
\hat{\Gamma}_{l,r}(m,k)=\frac{\hat{\Phi}_{l,r}(m,k)}{\sqrt{\hat{\Phi}_{l,l}(m,k)\hat{\Phi}_{r,r}(m,k)}}\, \text{,}
\end{equation}
%%%%END EQUATION%%%%
where $\hat{\Phi}_{x,y}(.)$ is the estimated cross-power spectrum for signals $x$ and $y$ and we use the short-hand index notation $l$ and $r$ to represent the left and right ear microphone signals $X_{l,r}(m,k)$, respectively. We estimate $\hat{\Phi}_{x,y}(.)$ recursively across time frames by $\hat{\Phi}_{x,y}(m,k)=\lambda\hat{\Phi}_{x,y}(m-1,k)+(1-\lambda)X_{x}(m,k) X_{y}^{*}(m,k)$, where $\lambda$ is a smoothing factor between 0 and 1 and $\ast$ indicates the complex conjugate operation.
We propose a new heuristic and parameterized formula for a DOA-dependent, binaural CDR estimator, $\widehat{CDR}(m,k)$, that is derived solely from $\hat{\Gamma}_{l,r}(m,k)$. For brevity and clarity in specifying the new CDR estimator, we omit the time and frequency indices. The new CDR estimator is given by:
%%%%% EQUATION %%%%%
\begin{equation}\label{eq:10}
\widehat{CDR}(\hat{\Gamma}_{l,r},S) = \Re\left\{\frac{{\exp}(\hat{\Gamma}_{l,r} + \cos(\arg(\hat{\Gamma}_{l,r})-(\pi/2)\arg(\hat{\Gamma}_{l,r})))}{\sqrt{\hat{\Gamma}_{l,r}+(S)\ln(\hat{\Gamma}_{l,r})+\cos(\arg(\hat{\Gamma}_{l,r})+\pi)}}\right\}\, \text{,}
\end{equation}
%%%%END EQUATION%%%%
where $\arg(\cdot)$ and $\Re\{\cdot\}$ refer to the phase and real part of a complex number, respectively, and $S$ is an adjustable positive real-valued parameter.  
%%%%%%%%%%%         FIGURE          %%%%%%%%%%%%
\begin{SCfigure}[][t]
\centering
\caption{The phasor response (geometrical locus) for the new CDR estimator ($ A= 1$)  is graphed for several coefficients \textit{S} in (a); the corresponding estimated gain responses are also shown in (b). A graph of the phasor response for the new CDR estimator is shown for $S=1$ and $0.1\leq A \leq1$ in (c) and the corresponding gain responses are shown in (d).}
\label{bcnr_locus}
\includegraphics[width=3.5in]{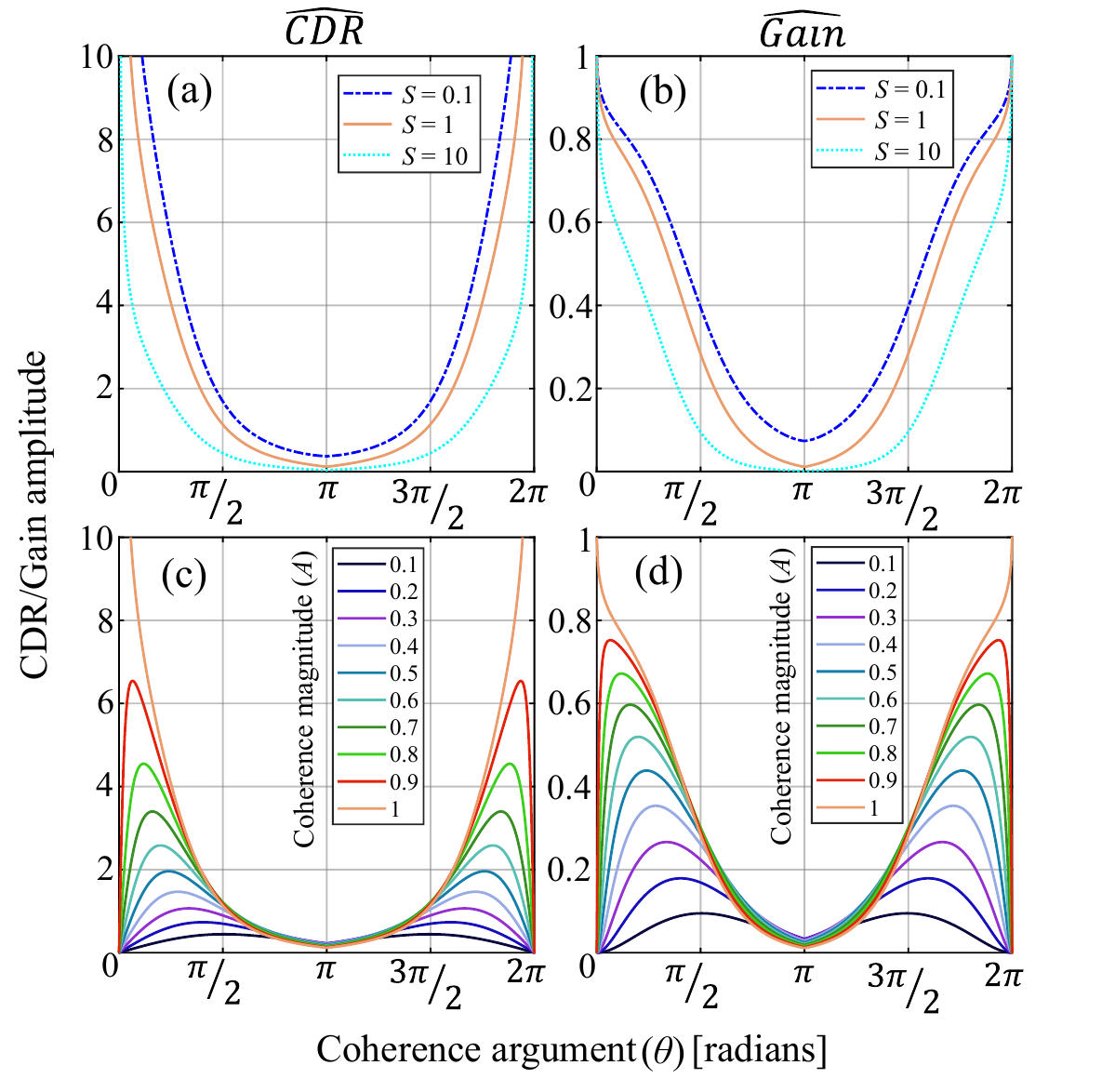}
\end{SCfigure}
%%%%%%%%%%%       END   FIGURE       %%%%%%%%%%
To clarify this formula, we consider Fig.\ref{bcnr_locus}a, which depicts the new CDR estimator as a function of the complex spatial coherence vector and several values of $S$. The complex spatial coherence vector can be represented by $\hat{\Gamma}_{l,r}=Ae^{j\theta}$, with amplitude, $0<A\leq 1$, and phase, $ 0\leq\theta\leq 2\pi$. From Fig.\ref{bcnr_locus}a we observe that  $\widehat{CDR}(\hat{\Gamma}_{l,r},S)$ is a U-shape (parabolic-like) function of the phase of the spatial coherence vector that is mirror-symmetric about $\theta = \pi$. We consider also Fig.\ref{bcnr_locus}c which shows the new CDR estimator for several values of the amplitude $A$ for $S = 1$. In general terms, the CDR estimator increases as $\theta$ approaches $0$ and as $A \rightarrow 1$; in other words, the geometrical slope of the $\widehat{CDR}(\hat{\Gamma}_{l,r},S)$ graph scales with the magnitude and phase of the spatial noise field coherence vectors. The variation in the geometrical pattern of $\widehat{CDR}(\hat{\Gamma}_{l,r},S)$ with changes in the $S$-value modifies the spatial directivity of the microphone system, i.e., higher $S$-values can reduce $\widehat{CDR}(\hat{\Gamma}_{l,r},S)$ as $\theta \rightarrow \pi$ which is equivalent to increasing the suppression of sound as it becomes incident from the side. The adjustability of the estimated CDR patterns may be useful for matching their values with the actual noise diffuseness in a specific frequency band. Depending upon the value of $A$, \emph{e.g.} $A < 1$, one notices that the CDR estimator demonstrates a peak that progressively moves away from $\theta = 0$ as $A$ decreases. More specifically, the formula has been empirically designed so that for $A < 0.94$, the CDR estimator goes to $0$ as $\theta \rightarrow 0$ with a faster rate compared to $ 0.94 < A < 1$. This behavior has been explicitly designed into the CDR estimator in order to suppress coherent noise that might arise at lower frequencies in a highly diffuse noise field, such as that related to the late reverberation of a room~\cite{jeub2011blind}. On the other hand, when $\theta \rightarrow \pi$, the observed geometrical pattern (see Fig.\ref{bcnr_locus}c) reduces the non-broadside PSD of the noise field in the binaural signals, which may be useful for preserving early source reflections and assisting with source localization based on interaural intensity differences.           

%%%%%%%%%%%         FIGURE          %%%%%%%%%%%%
\begin{figure*}[t]
\centering
\includegraphics[width=\textwidth]{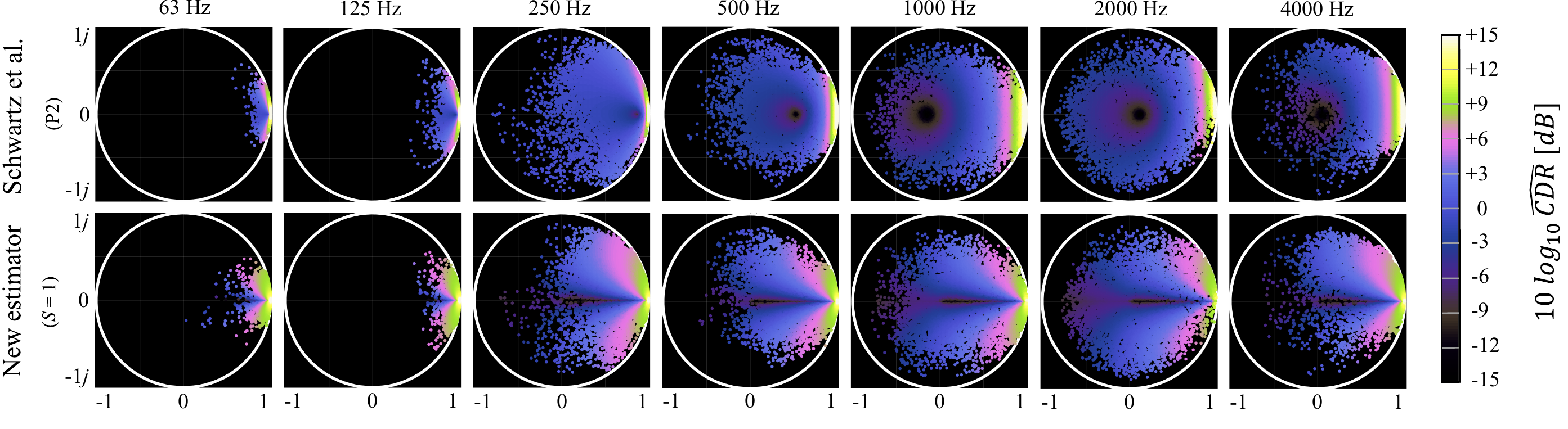}
\caption{The coherent-to-diffuse power ratio (CDR) is plotted as a function of the complex spatial coherence function, $\hat{\Gamma}_{l,r}(m,k)$, for the Schwartz \emph{et al.} P2 (Propose 2 in \cite{schwarz2015coherent}, TDOA=0) estimator and for the new estimator for $S=1$. The CDR levels for each frequency bin contains 7500 coherence vectors calculated for 60~s speech convolved with the broadside BRIRs ($d_{mic}$=17~m in the lecture room (AIR database) \cite{jeub2009binaural}.}
\label{cdr_vs_gammax_hot}
\end{figure*}
%%%%%%%%%%%       END   FIGURE       %%%%%%%%%%
To examine the relationship between of the new CDR estimator ($S=1$), $\widehat{CDR}(\hat{\Gamma}_{l,r},S)$, and the estimated, complex-valued spatial coherence function, $\hat{\Gamma}_{l,r}$ consider Fig.~\ref{cdr_vs_gammax_hot} that depicts the estimated CDR levels for a speech signal in a lecture room based on the position of spatial coherence vectors on the complex plane compared with the 'Propose~2' (P2) CDR estimator by Schwartz et al.~\cite{schwarz2015coherent}. Observe that the coherence vectors are more dispersed for higher frequencies but more concentrated around the positive real axis for lower frequencies. This phenomenon shows that low-frequency signals have higher correlations because of their comparatively long wavelengths concerning head size and microphone spacing.
The contrast between the two CDR estimators suggests that the new estimator may offer more reliable and precise estimation of CDR across frequency for a broadside signal located in front. For example, consider that the P2 CDR estimator shows a significant abrupt drop in the estimated CDR level as frequency decreases below 500~Hz, i.e.,  it will likely underestimate the low-frequency incident sound, while the new CDR estimator shows a more unbiased response across all frequencies. Significantly, one may also observe a spatial notch along the real axis for $\hat{\Gamma}_{l,r} < 0.94$ corresponding to the decreasing peak height shown in Fig.\ref{bcnr_locus}c. This spatial notch is intended to de-emphasize the diffuse noise signals while preserving the coherent direct signal.

\subsection{Binaural spectral enhancement}
\label{sec:binaural_enhancement}
\begin{SCfigure}[][t]
\centering
\includegraphics[width=3.5in]{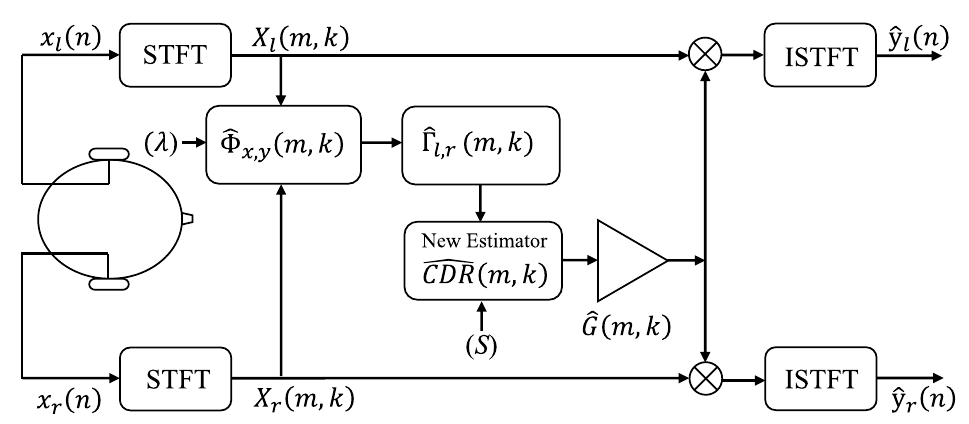}
\caption{Block diagram of the proposed binaural signal processing method. }
\label{block_diagram}
\end{SCfigure}

The application of the new CDR estimator for binaural noise and reverberation suppression is tested and investigated using methods like those proposed by \cite{jeub2010model}, as indicated in the block diagram in Fig.~\ref{block_diagram}. Observe that, the spatial CDR is first estimated as described in Section~\ref{sec:cdr_calculation} and a gain function, $\hat{G}(m,k)$, is then derived from the CDR estimate in the time-frequency domain as follows:
%%%%% EQUATION %%%%%
\begin{equation}\label{eq:11}
\hat{G}(m,k)=max\left( G_{min},\left(1-\frac{\mu}{\widehat{CDR}(m,k)+1}\right)^2\right)\, \text{.}
\end{equation}
%%%%END EQUATION%%%%

%%%%%%%%%%%         FIGURE          %%%%%%%%%%%%
%%%%%%%%%%%       END   FIGURE       %%%%%%%%%%
The aforementioned coherence-based gain function is equivalent to the square of a Weiner filter where $G_{min}$ is the gain floor to reduce the musical artifacts, and $\mu$ is referred to as the over-subtraction factor \cite{schwarz2015coherent} and is commonly set to one. The gain function is applied equally to the left and right channels, preserving the spatial auditory cues of interaural time and level differences. The effect of the gain function is shown in Figs.~\ref{bcnr_locus}b~and~\ref{bcnr_locus}d and generally follows the functional form of the new CDR estimator. As shown in Fig.~\ref{bcnr_locus}b, smaller values of $S$ result in less spatial noise/reverberation suppression and wider spatial directivity. In contrast, the larger values of $S$ increase the suppression of the noise/reverberation, provide narrower spatial filtering and may also increase audible artifacts. Furthermore, the square of the Wiener filter has been selected as the gain function since empirical testing has shown that it performs well with the new CDR estimator. i.e., it produces less audible artifacts and higher background noise suppression than a gain function based on the spectral magnitude subtraction as suggested by~\cite{schwarz2015coherent}.

\subsection{Binaural room simulations}
\label{binaural simulation}
Three state-of-the-art coherence-based dereverberation algorithms are compared with the proposed speech enhancement algorithm discussed in Section~\ref{sec:binaural_enhancement} in a binaural format. The counterpart CDR estimators used in this study are: (1) the DOA dependent CDR estimator ‘Propose 2’ (P2) in Schwartz \emph{et al.}~\cite{schwarz2015coherent}; (2) the DOA independent CDR estimator ‘Propose 3’ (P3) in Schwartz \emph{et al.}~\cite{schwarz2015coherent}; and (3) the effective rank-based DOA independent CDR estimator proposed by Löllmann \emph{et al.}~ \cite{lollmann2021effective}.
 All signal processing algorithms use a common  16kHz sampling rate, an FFT size of 512, a window length of 1024, and a hop size of 128 in MATLAB implementation. The gain function for the new CDR estimator was computed as described in Section~\ref{sec:binaural_enhancement} and for the counterpart algorithms the applied gain function is the spectral magnitude subtraction as described in   \cite{schwarz2015coherent,lollmann2021effective}, with $\mu=1$ and $G_{\text{min}} = 0.1$ for all algorithms. For the CDR estimators P2 and P3, the spatial coherence model for the diffuse noise is given as:$~\tilde{\Gamma}_{x,y}^{\text{diff}}(3D)=\sin{(2\pi f d_\text{mic} /c)}/(2\pi f d_{\text{mic}} /c)$, where $d_\text{mic}$ is the distance between two conchal microphones and $c$ is the speed of sound. For the CDR estimator P2, the spatial coherence for the broadside direct signal is taken as $1$ (real valued), while the CDR estimator P3 does not require an estimate of the spatial coherence of the direct sound \cite{schwarz2015coherent}. For the three counterparts, the smoothing factor $\lambda$ was set according to the relevant reference publication (P2 and P3: $\lambda = 0.68$; Löllmann: $\lambda = 0.8$). For the new CDR estimator, we chose $\lambda = 0.72$.\par
 
For the room simulation, we used a set of binaural HRIRs recorded by the conchal microphones of a generic in-the-ear hearable for a male subject with large pinna provided by the database described in~\cite{denk2020hearpiece}. The database HRIRs were then evenly interpolated for 642 directions on the surface of an imaginary sphere and used as input for the room simulator MCROOMSIM~\cite{wabnitz2010room} in order to obtain a set of BRIRs corresponding to a shoebox large room (20m x 16m x 5m) with 4 different reverberation times (0.3, 0.5, 1 and 2)s. The simulations were conducted with the listener positioned in the center of the room (ear level at 1.6~m). A target talker directional source is positioned in front of the listener at 0.5~m distance, and the subject is surrounded by a combination of one near-field time-reversed directional female speech masker located on the right and four far-field evenly distributed female speech maskers. The 34~s female utterances were derived from HARVARD speech corpus \cite{Philippa-2019}. In order to simulate a more realistic environment, a low-pass-filtered white noise (cutoff frequency 400 Hz) was mixed with the five masker signals. The relative signal levels used for the target, maskers, and low-pass filtered noise signals were varied and specified as a triplet of numbers (0, -6, -10) dB, (0, 0, -10) dB and (-6, 0, 0) dB, respectively. The process above was repeated for four different reverberation times, set by changing the room acoustic absorption settings in MCROOMSIM.

\subsection{Broadband low-latency real-time framework}
\label{real-time}
\begin{SCfigure}[][t]
    \centering
    \includegraphics[width=3in]{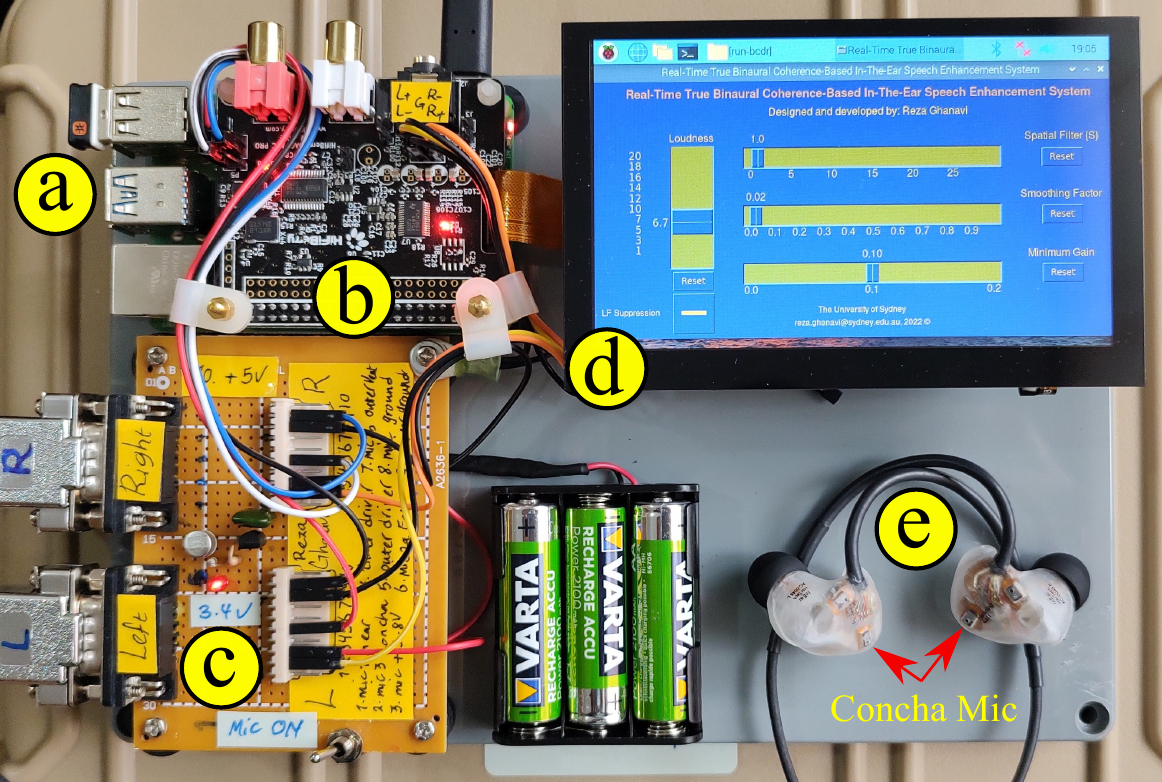}
    \caption{Real-time low-latency prototype of the new CDR-based true binaural speech enhancement system. Raspberry Pi 4B (a), HiFiBerry DAC plus ADC Pro (b), hearable interface (C), online user interface (d) and binaural in-the-ear earpieces  ~\cite{denk2020hearpiece} (e). }
    \label{rt-prototype}
\end{SCfigure}

Fig.\ref{rt-prototype} shows the actual Raspberry Pi-based embedded system \cite{carvalho2019raspberry} prototype adapted for high-quality and low-latency online implementation of the described new algorithm in Python \cite{de2020programming}. The recorded binaural time signals (32kHz sample rate) are buffered (window size 512) using 50\% overlap. The selection of larger window sizes enables more accurate short-time signal power estimation at lower frequencies and has shown fewer artifacts in a time-variant system. Furthermore, the number of FFT points has been doubled (FFT size = 1024) to improve the quality of spectral enhancement processing in the frequency domain. Using the Hanning window, the signal is then reconstructed via the weighted overlap-add (WOLA) technique \cite{crochiere1980weighted}. The output buffer is filled by the second half of the previous segment in time and the first half of the current segment to reduce real-time latency by half. In this case, the total acoustic latency in this system is measured to be about 9~ms, which is comparable with the average latency in a high-quality hearing aid system \cite{alexander2019hearing}. The user interface for this system enables online adjustment of the S-parameter as well as other parameters.

\section{\uppercase{RESULTS}}
\label{results}
\subsection{Objective Speech Enhancement Performance}
\label{objective-results}
The performance of the new algorithm for speech dereverberation and denoising is compared with other state of the art algorithms as mentioned earlier using two intrusive methods: perceptual evaluation of speech quality (PESQ) \cite{rix2001perceptual} and cepstrum distance (CD); and two non-intrusive methods:  speech-to-reverberation modulation energy ratio (SRMR) and word error rate (WER) calculated for an automatic speech recognition (ASR) system. The narrow-band results for PESQ and CD are averaged across the two binaural enhanced signals, while the SRMR and WER data are derived based on a monaural mix-down of the two full-band binaural signals.\par
%%%%%%%%%%%         FIGURE          %%%%%%%%%%%%

\begin{figure}[h]
\centering
\includegraphics[width=5in]{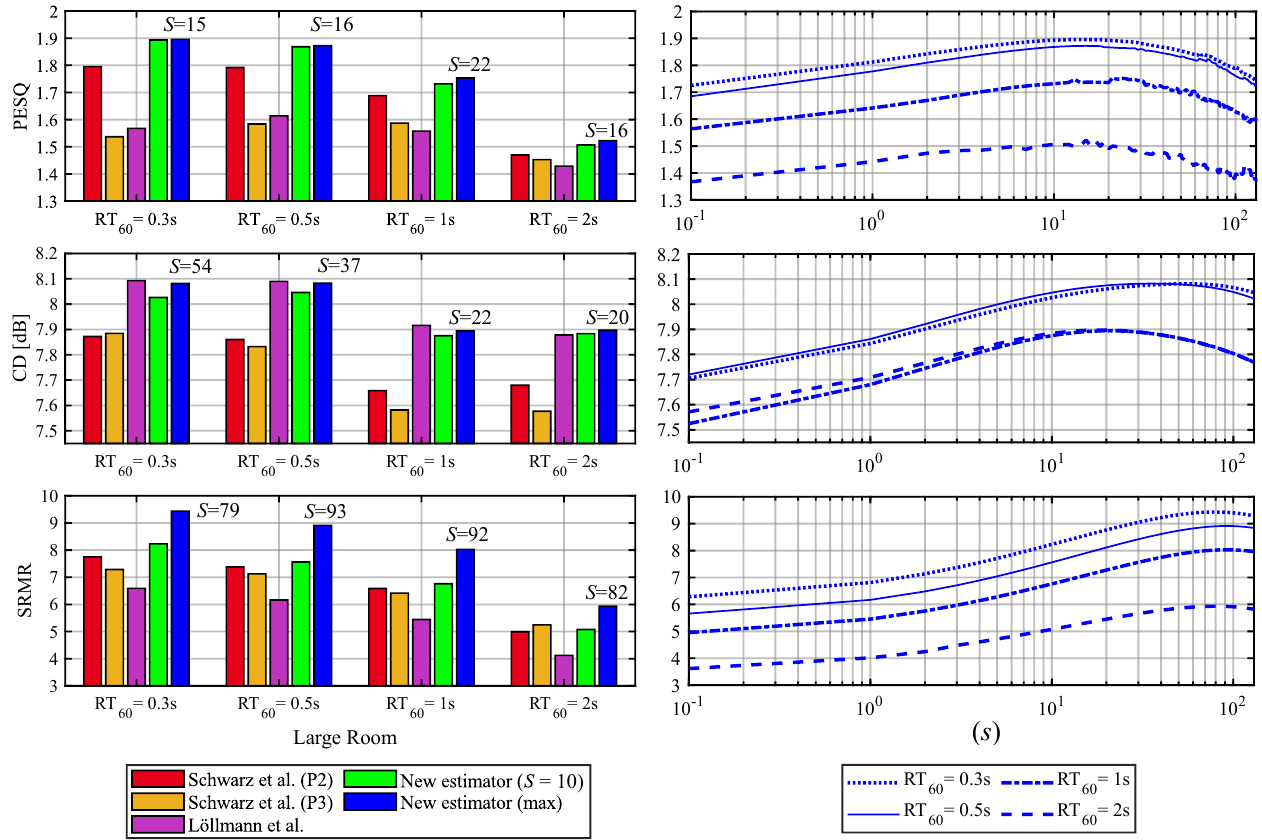}
\caption{Objective performance results of the new reverberation suppression algorithm are compared with the performance of the counterpart algorithms in the multi-talker scenarios simulated for a large room (left bar graphs) and different values of coefficient \textit{S} (right curves).}
\label{R7_pesq_cd_srmr_cpp_modified}
\end{figure}

%%%%%%%%%%%       END   FIGURE       %%%%%%%%%%
In Fig.\ref{R7_pesq_cd_srmr_cpp_modified}, the calculated objective values are averaged over the results derived for several signal-to-masker/noise ratios (refer to Section \ref{binaural simulation}). Observe first that the $S$-values obtaining the best results for the newly proposed estimator vary across the various speech quality measures (to a larger extent) and also across the different reverberation conditions (to a lesser extent), with the optimal $S$-value for a given acoustic environment as the performance measure changes from PESQ to CD to SRMR.
The performance variations depending on the $S$-parameter demonstrate that the three measures examine various aspects of the direct sound and background noise quality. The new CDR estimator with the optimal $S$-value generally improves the PESQ values compared with the other CDR estimators, e.g., for $S=10$, the sound quality of the new algorithm generally outperforms the counterpart algorithms for all of the reverberant conditions. Interestingly, a significant increase in the SRMR performance values is found for the new CDR estimator with optimal $S$-values compared with the other CDR estimators. However, the $S$-value must be increased significantly to obtain these results, i.e., higher $S$ values result in a more direct-to-reverberant ratio (DRR). However, it can be observed that in a given room, for $S > 100$, the spatial gains for the direct signal can be declined significantly due to change in the shape and slope of the new CDR function curves (see Fig.\ref{bcnr_locus}a). In addition, the increase in $S$ values can make more modifications to the background noise spectrum that may explain the slight increments in the CD values. The varying $S$-values obtaining optimal performance across the three speech enhancement measures indicate that there are different and likely conflicting requirements for optimizing speech enhancement performance based on CDR, depending on the significance given to a particular speech enhancement measure.\par
%%%%%%%%% TABLE  2  %%%%%%%%%%%%
\begingroup
\setlength{\tabcolsep}{1pt} % Default value: 6pt
\renewcommand{\arraystretch}{1} % Default value: 1
\begin{table}[!ht]
\caption{Averaged word error rate (WER)}
\begin{threeparttable}
\vspace*{3mm}
\label{table1}
\small
\setlength\tabcolsep{10pt} % make LaTeX figure out intercolumn spacing
%\begin{center}
\begin{tabular*}{\textwidth}{@{\extracolsep{\fill}} ccccccccc}

\toprule
&& \multicolumn{2}{c}{Schwarz et al.} & \multicolumn{1}{c}{Löllmann et al.}& \multicolumn{4}{c}{New estimator} \\ 
\cmidrule{3-4}     
\cmidrule{6-9}
  
              RT60(s)&Unprocessed&(P2)    &(P3)	 	  && S = 0.1 & S = 1	& S = 3 & S = 10\\
\midrule
                 0.3	&55.67	&45.33	&62.67	&60.33	&\textbf{45}	     &48.33	   &50.67	&46\\
                 0.5	&53.33	&50.67	&57.67	&57.67	&50	     &48	   &\textbf{44}	    &46\\
                 1	    &55	    &52.33	&58.67	&53.67	&49.33   &47.33    &\textbf{47}	    &50.33\\
                 2	    &62.33	&\textbf{59.33}	&64.33	&61.33	&\textbf{59.33}	 &60.67	   &60	    &64.67\\

\midrule        
                   Average	&56.58	&51.92	&60.83	&58.25	&50.92	&51.08	&\textbf{50.42}	&51.75\\
\bottomrule
\end{tabular*}
\smallskip
\scriptsize
%\end{center}
\end{threeparttable}
\end{table}
%%%%%%%%% END of TABLE 2 %%%%%%%%%%%%%%%
The objective speech intelligibility was also estimated by calculating the word error rate (WER) of the automated speech recognition (ASR) algorithm for the processed and unprocessed speech. The ASR engine Deepspeech 0.9.3 was used \cite{hannun2014deep}. In this work, we used clean speech containing 34~s of female speech (100 words) from the HARVARD speech corpus that was $100\%$ recognizable by the pre-trained ASR engine. The average word error rates can thus be attributed to the acoustic condition and binaural sound processing systems.
Table 1 shows the WER results for the multi-talker scenarios. On average, the word error rate for the new CDR estimator enhancement algorithm with $S=3$ is lower compared with the other CDR-based algorithms. For the multi-talker scenario, the WER results indicate potential advantages to be found by tuning the $S$-value specifically for a given sound environment.

\subsection{Preliminary subjective evaluation}
\label{subjective-results}
The binaural psychoacoustic perception of the newly proposed algorithm compared with Schwartz et al., \cite{schwarz2015coherent} (P2 and P3) is evaluated through several informal listening tests in regular rooms and a large reverberant/noisy cafeteria using the real-time platform described in section \ref{real-time}. We are preparing to conduct a proper psychoacoustic experiment in the future. Here we report some anecdotal results. For all algorithms, the optimized online parameters $\lambda$ and $G_{min}$ are set to 0.02 and 0.1, respectively. In general terms, the perceived sound quality is compatible with the objective results discussed in Section~\ref{objective-results}; however, the informal listening tests have revealed that the new algorithm seem to significantly improve the spatial quality of the sound in a given environment compared with the the counterpart algorithms. For example, the perceived frontal near-field and far-field binaural intelligibility in the presence of several random distributed noise/masker sources is highly improved for the new CDR formula, while Schwartz et al., P2 has shown satisfactory results for only the near-field target and P3 has shown less enhancement for the target-to-masker ratio. Furthermore, the enhanced multi-talker and noisy spatial atmosphere reproduced by the new binaural algorithm was reported as perceived as more natural, robust and quiet compared with the two other algorithms. i.e., accurate source localization and externalization are preserved for the new CDR estimator resulting in improved listening comfort. The online adjustment of parameter $S$ has revealed that small changes in $S < 20 $ are perceivable and may be advantageous in adjusting the enhanced target speech quality and spatial perception in natural environments. The increment of the $S$ value for $S>10$ can produce minor audible artifacts due to a higher level of background noise modification.

\section{\uppercase{Conclusions}}
A new binaural, directional coherent-to-diffuse power ratio (CDR) estimator has been proposed for noise reduction and dereverberation in multi-talker reverberant and noisy environments. The CDR estimator relies only on the observed complex coherence between the binaural microphones and maximizes for a broadside target signal. The binaural application of the gain function and the compatibility of the new formula with the actual binaural noise field preserves spatial hearing cues. Furthermore, the new CDR estimator employs a variable $S$-parameter to provide adjustable coherence-based spatial filtering for different noise conditions.

The objective numerical evaluations show that varying the $S$-parameter enables the CDR estimate to improve speech quality and/or the direct-to-reverberant ratio. Adjusting the $S$-parameter enables a trade-off between signal quality, degree of dereverberation, and the spatial quality of the sound. The results suggest that the new CDR estimator may improve existing coherence-based methods for denoising and dereverberation. To this end, several informal listening tests have shown more advantages of the new method compared with two counterpart algorithms in terms of sound naturalness, accurate source localization and voice intelligibility.

\section*{\uppercase{Acknowledgements}}
The authors would like to thank Jorg Buchholz (Macquarie University) for his constructive comments and suggestions. This research is supported by an Australian Government Research Training Program (RTP) Scholarship.

%\hl{AO: I have not updated the reference style yet...}

\renewcommand{\refname}{\normalfont\selectfont\normalsize}
\noindent \section*{\uppercase{References}}
\vspace{-18pt}

\bibliography{./article_1_bib}

\begin{thebibliography}{10}

\bibitem{arons1992review}
Arons B.
\newblock A review of the cocktail party effect.
\newblock Journal of the American Voice I/O Society. 1992;12(7):35-50.

\bibitem{ebata2003spatial}
Ebata M.
\newblock Spatial unmasking and attention related to the cocktail party
  problem.
\newblock Acoustical Science and Technology. 2003;24(5):208-19.

\bibitem{parande2017study}
Parande PG, Thomas T.
\newblock A study of the cocktail party problem.
\newblock In: 2017 International Conference on Electrical and Computing
  Technologies and Applications (ICECTA). IEEE; 2017. p. 1-5.

\bibitem{qian2018past}
Qian Ym, Weng C, Chang Xk, Wang S, Yu D.
\newblock Past review, current progress, and challenges ahead on the cocktail
  party problem.
\newblock Frontiers of Information Technology \& Electronic Engineering.
  2018;19(1):40-63.

\bibitem{jeub2011blind}
Jeub M, Nelke C, Beaugeant C, Vary P.
\newblock Blind estimation of the coherent-to-diffuse energy ratio from noisy
  speech signals.
\newblock In: 2011 19th European Signal Processing Conference. IEEE; 2011. p.
  1347-51.

\bibitem{mccowan2003microphone}
McCowan IA, Bourlard H.
\newblock Microphone array post-filter based on noise field coherence.
\newblock IEEE Transactions on Speech and Audio Processing. 2003;11(6):709-16.

\bibitem{schwarz2015coherent}
Schwarz A, Kellermann W.
\newblock Coherent-to-diffuse power ratio estimation for dereverberation.
\newblock IEEE/ACM Transactions on Audio, Speech, and Language Processing.
  2015;23(6):1006-18.

\bibitem{thiergart2012signal}
Thiergart O, Del~Galdo G, Habets EA.
\newblock Signal-to-reverberant ratio estimation based on the complex spatial
  coherence between omnidirectional microphones.
\newblock In: 2012 IEEE International Conference on Acoustics, Speech and
  Signal Processing (ICASSP). IEEE; 2012. p. 309-12.

\bibitem{thiergart2012spatial}
Thiergart O, Del~Galdo G, Habets EA.
\newblock On the spatial coherence in mixed sound fields and its application to
  signal-to-diffuse ratio estimation.
\newblock The Journal of the Acoustical Society of America.
  2012;132(4):2337-46.

\bibitem{jeub2010binaural}
Jeub M, Vary P.
\newblock Binaural dereverberation based on a dual-channel wiener filter with
  optimized noise field coherence.
\newblock In: 2010 IEEE International Conference on Acoustics, Speech and
  Signal Processing. IEEE; 2010. p. 4710-3.

\bibitem{jeub2010model}
Jeub M, Schafer M, Esch T, Vary P.
\newblock Model-based dereverberation preserving binaural cues.
\newblock IEEE Transactions on Audio, Speech, and Language Processing.
  2010;18(7):1732-45.

\bibitem{westermann2013binaural}
Westermann A, Buchholz JM, Dau T.
\newblock Binaural dereverberation based on interaural coherence histograms.
\newblock The Journal of the Acoustical Society of America.
  2013;133(5):2767-77.

\bibitem{lollmann2021effective}
L{\"o}llmann HW, Brendel A, Kellermann W.
\newblock Effective Rank-Based Estimation of the Coherent-to-Diffuse Power
  Ratio.
\newblock In: ICASSP 2021-2021 IEEE International Conference on Acoustics,
  Speech and Signal Processing (ICASSP). IEEE; 2021. p. 955-9.

\bibitem{jeub2009binaural}
Jeub M, Schafer M, Vary P.
\newblock A binaural room impulse response database for the evaluation of
  dereverberation algorithms.
\newblock In: 2009 16th International Conference on Digital Signal Processing.
  IEEE; 2009. p. 1-5.

\bibitem{denk2020hearpiece}
Denk F, Kollmeier B.
\newblock The Hearpiece database of individual transfer functions of an openly
  available in-the-ear earpiece for hearing device research.
\newblock arXiv preprint arXiv:200406579. 2020.

\bibitem{wabnitz2010room}
Wabnitz A, Epain N, Jin C, Van~Schaik A.
\newblock Room acoustics simulation for multichannel microphone arrays.
\newblock In: Proceedings of the International Symposium on Room Acoustics.
  Citeseer; 2010. p. 1-6.

\bibitem{Philippa-2019}
Philippa D. HARVARD speech corpus - audio recording 2019.; 2019.
\newblock University of Salford.
\newblock Available from:
  \url{https://doi.org/10.17866/rd.salford.c.4437578.v1}.

\bibitem{carvalho2019raspberry}
Carvalho A, Machado C, Moraes F.
\newblock Raspberry Pi Performance Analysis in Real-Time Applications with the
  RT-Preempt Patch.
\newblock In: 2019 Latin American Robotics Symposium (LARS), 2019 Brazilian
  Symposium on Robotics (SBR) and 2019 Workshop on Robotics in Education (WRE).
  IEEE; 2019. p. 162-7.

\bibitem{de2020programming}
De~Pra Y, Fontana F.
\newblock Programming real-time sound in python.
\newblock Applied Sciences. 2020;10(12):4214.

\bibitem{crochiere1980weighted}
Crochiere R.
\newblock A weighted overlap-add method of short-time Fourier
  analysis/synthesis.
\newblock IEEE Transactions on Acoustics, Speech, and Signal Processing.
  1980;28(1):99-102.

\bibitem{alexander2019hearing}
Alexander J, et~al.
\newblock Hearing aid delay and current drain in modern digital devices.
\newblock Canadian Audiologist. 2019;3(4).

\bibitem{rix2001perceptual}
Rix AW, Beerends JG, Hollier MP, Hekstra AP.
\newblock Perceptual evaluation of speech quality (PESQ)-a new method for
  speech quality assessment of telephone networks and codecs.
\newblock In: 2001 IEEE international conference on acoustics, speech, and
  signal processing. Proceedings (Cat. No. 01CH37221). vol.~2. IEEE; 2001. p.
  749-52.

\bibitem{hannun2014deep}
Hannun A, Case C, Casper J, Catanzaro B, Diamos G, Elsen E, et~al.
\newblock Deep speech: Scaling up end-to-end speech recognition.
\newblock arXiv preprint arXiv:14125567. 2014.

\end{thebibliography}

\end{document}